\documentclass[twocolumn]{svjour3}          % twocolumn
\smartqed  																	% flush right qed marks, e.g. at end of proof
\usepackage{graphicx}												% include figure files
\usepackage{mathptmx}      									% use Times fonts if available on your TeX system
\usepackage{color}													% Allows for colored text
\usepackage{hyperref}												% Allows for hyperlinked references.
\usepackage[misc,geometry]{ifsym}  					% Used to get the corresponding author symbol.

\journalname{arXiv}
\begin{document}

\title{Laser ablation production of Ba, Ca, Dy, Er, La, Lu, and Yb ions%\thanks{Grants or other notes
			}

\author{S. Olmschenk \and
        P. Becker %etc.
			 }

\institute{S. Olmschenk (\Letter) $\cdot$ P. Becker \at
              Denison University, 100 West College Street, Granville, OH 43023 \\
							Tel.: 740.587.8661\\
              Fax: 740.587.6240\\
              \email{steven.olmschenk@denison.edu}
					}

\date{Received: 2016.11.29 / Accepted: 2017.02.22 / Published: 2017.03.17}
\maketitle

%%%%%%%%%%%%%%%%%%%%%%%%%%%%%%%%%%%%%%%%%%
\begin{abstract}
We use a pulsed nitrogen laser to produce atomic ions by laser ablation, measuring the relative ion yield for several elements, including some that have only recently been proposed for use in cold trapped ion experiments.  For barium, we monitor the ion yield as a function of the number of applied ablation pulses for different substrates.  We also investigate the ion production as a function of the pulse energy, and the efficiency of loading an ion trap as a function of radiofrequency voltage.
\end{abstract}

%%%%%%%%%%%%%%%%%%%%%%%%%%%%%%%%%%%%%%%%%%
\section{Introduction}
\label{intro}
Trapped atomic ions are used in a range of experiments and applications, including atomic clocks~\cite{ludlow:optical_atomic_clocks}, cold chemistry~\cite{harter:cold_atom-ion_expts}, precision measurements of fundamental physics~\cite{karr:precision_measurements_ions,orzel:amo_new_physics}, and quantum information~\cite{monroe:scaling_ion_trap_qip}.  The long-term goals of these experiments have led to efforts to miniaturize the system to meet application specifications and scalability~\cite{schwindt:miniaturized_ion_clock}, and in some cases aims for cryogenic ion trap systems~\cite{demotte:expt_design_ion_squid,brandl:cryogenic_setup_ion_qc}.  Efforts using unconventional or multiple ion species put additional demands on the system.  These directions may require different methods for initial loading of the ion trap.

The standard method for loading an ion trap consists of producing a flux of neutral atoms, and then ionizing these atoms within the trapping volume.  A neutral flux of atoms is commonly produced by resistively heating a sample, and these atoms are then ionized at the trap using either electron bombardment~\cite{dehmelt:iontrap} or photoionization~\cite{kjaergaard:photoionization_mg_ca,gulde:photoionization_calcium,lucas:photoionization_calcium,tanaka:photoionization_rare_calcium,deslauriers:photoionization,brownnutt:photoionization_strontium,steele:photoionization_barium,wang:photoionization_barium_led,johanning:photoionization_ytterbium,leschhorn:photoionization_barium}.  While electron bombardment is applicable to any atomic species, it can reduce the lifetime of the trapped ions by contaminating the vacuum, and the accumulation of charge on insulators can result in additional micromotion or even destabilize the trap.  Although photoionization can avoid the issues related to electron bombardment, there can be significant overhead associated with the additional lasers, which can become substantial when working with multiple elements, atoms requiring unconventional wavelengths, or multiply-charged ions.  Finally, producing neutral atom flux with a standard resistively-heated source can be difficult when working with elements that aggressively react in air~\cite{devoe:ba_heating,graham:barium_surface_trap}, refractory elements that require high temperatures, or where miniaturization or cryogenic designs may prohibit the use of these sources.  Given the limitations of standard loading procedures, other methods for loading ion traps are being actively pursued~\cite{cetina:mot_loading_ion_trap,sage:mot_loading_ion_trap}.

Laser ablation of a sample offers an alternative method for loading ion traps.  In general, a laser pulse with sufficient peak fluence incident on a sample can vaporize a fraction of the material, producing atoms, ions, molecules, and clusters~\cite{russo:laser_ablation,willmott:pulsed_laser_deposition}.  Thus, for ion trap experiments ablation can be used as a source of neutral atom flux~\cite{hendricks:ablation,sheridan:all_optical_ion_trap_loading}, or can directly produce ions~\cite{knight:ablation,gill:laser_ablation_ion_trap_ms,matsuo:ablation,gill:resonant_laser_ablation,hashimoto:ablation,leibrandt:ablation,kwapien:ablation,schmitz:laser_ablation_ion_trap_atmosphere,zimmermann:ablation_loading}, including multiply-charged species~\cite{kwong:ablation,campbell:thorium_crystals}. In reference~\cite{zimmermann:ablation_loading}, Zimmermann \textit{et al.} demonstrated the use of a compact nitrogen laser for producing ions by laser ablation, including some multiply-charged species.  Here we use a similar setup to measure the ion yield from several elements, including lanthanum, erbium, and dysprosium, which have only recently been proposed for use in cold trapped ion experiments~\cite{olmschenk:lanthanum_proposal,lepers:rare-earth_ion_cooling}.  We then focus on barium ion production, investigating the ion yield for different samples as a function of the number of ablation pulses applied, the ion yield as a function of the pulse energy, and the efficiency of loading an ion trap as a function of radiofrequency voltage.

%%%%%%%%%%%%%%%%%%%%%%%%%%%%%%%%%%%%%%%%%%
\section{Ion Production in a TOF Mass Spectrometer}
\label{sec:tofms}
\begin{figure}
  \includegraphics[width=1.0\columnwidth,keepaspectratio]{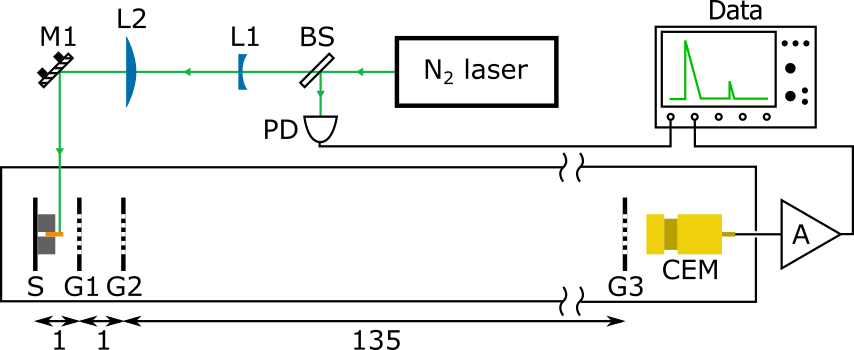}
\caption{Experimental TOF mass spectrometer setup (top view, schematic, not to scale).  Pulses from the nitrogen laser are focused onto the sample in the vacuum chamber.  A fraction of the pulse is directed to a photodiode to trigger the scope.  (Not shown:  a red diode laser beam is incident on the other port of the beam sampler, and is used for aligning the nitrogen laser pulses on the sample.)  BS is beam sampler; PD is photodiode; L1 is 1-inch diameter, $f=-100$ mm plano-concave lens; L2 is 2-inch diameter, $f=100$ mm plano-convex lens; M1 is mirror in a piezo-actuated mount.  The distance between L1 and L2 is about 16 cm; the path length from L2 to the sample is about 15 cm.  The sample is typically held at about 805 V, the first grid at about 730 V, and the second and third grids at ground.  S is sample plate; G1 is first grid; G2 is second grid; G3 is third grid; CEM is channeltron electron multiplier; A is amplifier.  Distances are given in cm.}
\label{fig:tof_setup}       % Give a unique label
\end{figure}

Ions are produced by laser ablation of a sample with pulses from a commercial nitrogen laser (Stanford Research Systems, NL100).  The pulses from this laser are specified by the manufacturer to have a wavelength of about 337 nm, maximum energy of about 170 $\mu$J, duration of about 3.5 ns, and repetition rates up to 20 Hz.  To increase the laser fluence for ablation, the pulses are focused onto the sample using a pair of lenses (L1 and L2), as shown in Fig.~\ref{fig:tof_setup}.  We measure the resulting spot size by attenuating the pulses, temporarily removing the mirror (M1), and placing a camera (Point Grey, FL3-U3-13S2M-CS) at the laser focus.  Pulsing the laser near its maximum repetition rate allows us to image and measure the spot size, resulting in elliptical beam waists of approximately 280 $\mu$m and 50 $\mu$m~\cite{gaussian_beam_note}.  When incident on the ablation target, we estimate the focal point to be within 1 mm of the sample, optimized by adjusting the position of the lenses to maximize the ion yield.  A beam sampler (BS) directs a fraction of each pulse to a photodiode (PD) to trigger the oscilloscope for data acquisition.  A mirror (M1) in a piezo-actuated mount (Newport, AG-M100N) can be used to sweep the pulses in each run through a location grid on the sample, but in practice this was found to be unnecessary for most samples.  In all cases, pulses are incident approximately normal to the sample.  %Might need additional details here?

We analyze the ions produced by ablation using a custom time-of-flight (TOF) mass spectrometer based on the Wiley-McLaren design~\cite{wiley:tof_mass_spect}, as shown in Fig.~\ref{fig:tof_setup}.  The sample (target) is mounted between two stainless steel blocks, which are mechanically and electrically attached to a stainless steel plate (S) held under vacuum at about $10^{-7}$ torr (pressure limited by chamber design and vacuum preparation, not the ablation process).  This sample plate is held at about 805 V, and the subsequent grid (G1) is held at about 730 V, directing ions produced at the sample toward the remainder of the mass spectrometer.  The second grid (G2) is held at ground, and the potential difference between G1 and G2 provides most of the acceleration of the ions.  After passing through G2, ions travel through a nominally electric field-free drift region for mass separation, before reaching the third grid (G3), which is also held at ground~\cite{wire_mesh_note}.  After passing through G3, the ions are detected by a channeltron electron multiplier (CEM; Photonis, Magnum 5900), which has a large negative potential applied to it (-2000 V).

\begin{figure}
  \includegraphics[width=1.0\columnwidth,keepaspectratio]{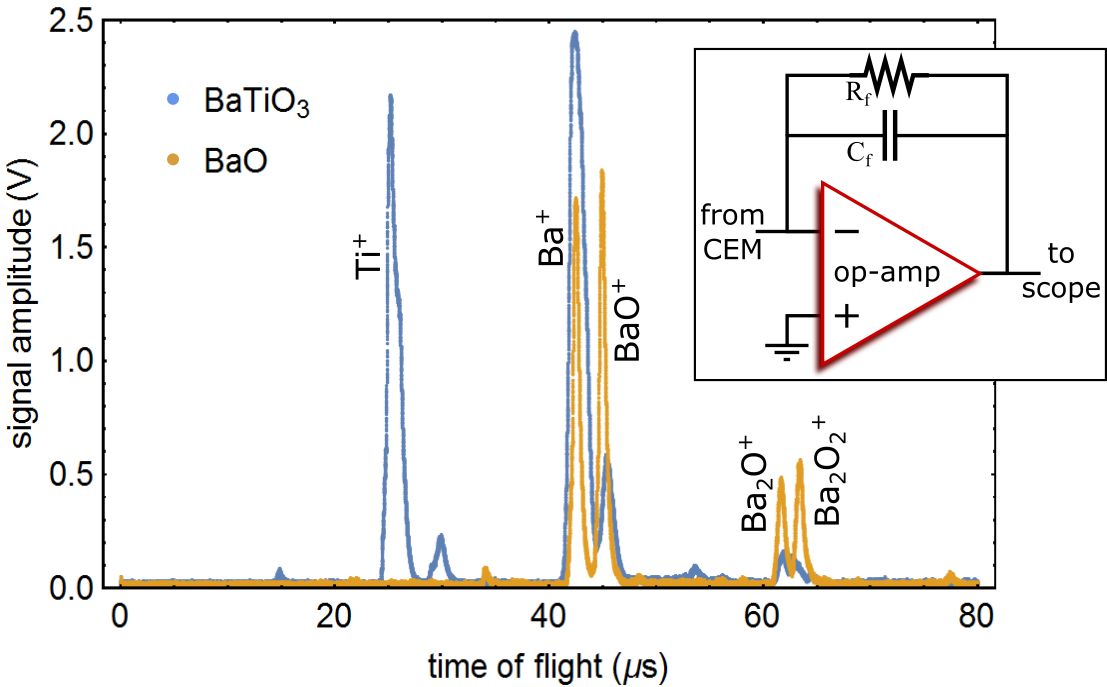}
\caption{Example signal acquisition (scan) following laser ablation of BaO and BaTiO${}_3$ targets, each averaged over 25 TOF spectra.  Major peaks are identified based on the time of flight and mass spectrometer grid potentials.  \textit{Inset}: Charge-sensitive amplifier schematic, used to amplify the signal from the CEM.  An AD8033 op-amp is used here, and feedback resistor (R${}_{\mbox{f}}=10$ k$\Omega$) and capacitor (C${}_{\mbox{f}}=30$ pF) values result in a signal decay time of about 0.3 $\mu$s.}
\label{fig:csa_scan}       % Give a unique label
\end{figure}

The signal from the CEM is amplified by a basic charge-sensitive amplifier, shown in the inset of Fig.~\ref{fig:csa_scan}.  The components in the circuit are chosen to provide good signal-to-noise, while keeping the decay time of the signal short compared to the expected flight time between ions with different charge-to-mass ratios.  Data for BaO and BaTiO${}_3$ targets, each averaged over 25 TOF spectra (25 ablation pulses), is shown in Fig.~\ref{fig:csa_scan} with major peaks identified.

The relative ion yield is determined for several elements, as shown in Table~\ref{tab:ion_yields}, where the values given are with respect to the the largest ion yield (Ca); for reference, the Ca TOF peak reaches about 6.9 V on the oscilloscope.  For each element, the data from four scans is analyzed, where each scan is an average of 25 TOF spectra.  We take the average peak height consistent with the charge-to-mass ratio of the investigated ion as the ion yield~\cite{detection_efficiency_note}.  In each case, approximately 100 ablation pulses are applied to the sample before the four scans that constitute the data analyzed.  (It is found that the first several pulses yield additional ablation products not seen in later scans, and that subsequent TOF spectra are more consistent.  For some elements, peaks consistent with atomic and molecular nitrogen, oxygen, etc., remain in subsequent TOF spectra.)  Note that the yield presented in Table~\ref{tab:ion_yields} for barium (Ba) is with a BaTiO${}_3$ target, as discussed below, whereas all the others are for a pure ($\geq99\%$) element target.  Additionally, while the scans for barium, dysprosium, erbium, lanthanum, and lutetium show minor peaks consistent with doubly-ionized atoms, these signals are not unambiguously identified due to the resolution of our spectrometer, so only singly-ionized yields are tabulated.

\begin{table}
\centering	% center the table in the column
	\caption{Relative yield of each ion by laser ablation, with respect to the largest yield (Ca).  Note that the yield presented for Ba is with a BaTiO${}_3$ target, whereas all the others are for a pure element target.  Ion yield is determined by the height of the peak detected in the TOF mass spectrometer consistent with the charge-to-mass ratio of the species analyzed.  For each ion listed, four scans were analyzed, where each scan is the average of 25 TOF spectra (25 ablation pulses), and the value in parentheses is the standard deviation of the four scans.}
	\label{tab:ion_yields}	% Give a unique label
	\begin{tabular}{lr}
		\hline%\noalign{\smallskip}
		Ion & Relative Yield \\
		\hline%\noalign{\smallskip}
		\hline
		Ca & 1.~~ (0.02) \\
		Ba & 0.37 (0.01) \\
		Dy & 0.56 (0.02) \\
		Er & 0.46 (0.01) \\
		La & 0.48 (0.03) \\
		Lu & 0.52 (0.03) \\
		Yb & 0.51 (0.01) \\
		\hline
	\end{tabular}
\end{table}

\begin{figure}
  \includegraphics[width=1.0\columnwidth,keepaspectratio]{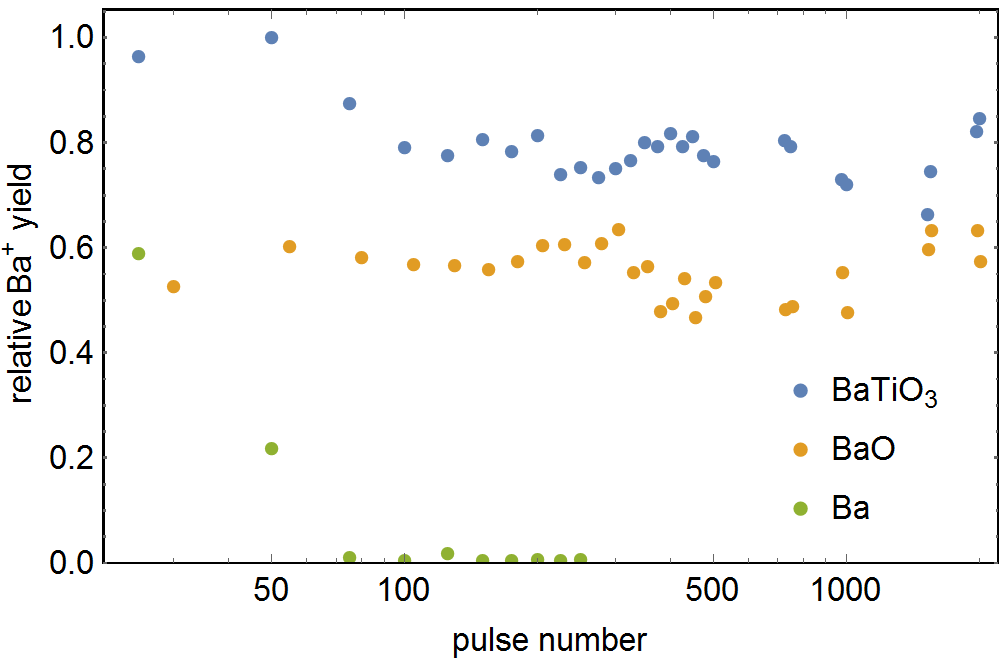}
\caption{Barium ion yield versus pulse number.  The Ba${}^+$ yield is roughly constant over at least 2000 ablation pulses on a single location for both BaO and BaTiO${}_3$ targets, with the BaTiO${}_3$ target yield consistently higher.  Conversely, the Ba${}^+$ yield from the Ba target ($\geq99\%$ pure barium, exposed to atmosphere for approximately 30 minutes) quickly decreases as a function of ablation pulse number.  In this case, the data for Ba was obtained by sweeping the ablation laser position across a 5x5 location grid on the sample; due to the rapid decrease in ion yield, we could not obtain a satisfactory average of 25 TOF spectra at a single location.}
\label{fig:barium_ions_vs_pulse_number}       % Give a unique label
\end{figure}

We verify that the ion yield for each element remains roughly constant for application of at least 500 ablation pulses focused onto a single location on the sample.  The exception to this is elemental barium, exposed to atmosphere for about 30 minutes.  As shown in Fig.~\ref{fig:barium_ions_vs_pulse_number}, the ion yield for this target quickly decreases, even when the ablation pulses are swept across a 5x5 location grid in each scan.  The unreliability of the elemental barium target led us to investigate alternative ablation targets, BaO and BaTiO${}_3$.  Both of these targets produce a roughly constant yield of Ba${}^+$ for at least 2000 ablation pulses at a single location on the sample (Fig.~\ref{fig:barium_ions_vs_pulse_number}).  Here, the BaO is a sample of pure ($\geq99\%$) Ba exposed to atmosphere for more than a year.  The ability to produce ions from an oxidized target may be particularly useful for samples that react aggressively in air, such as barium, since it is also more difficult to construct conventional atomic sources~\cite{devoe:ba_heating,graham:barium_surface_trap}.

\begin{figure}
  \includegraphics[width=1.0\columnwidth,keepaspectratio]{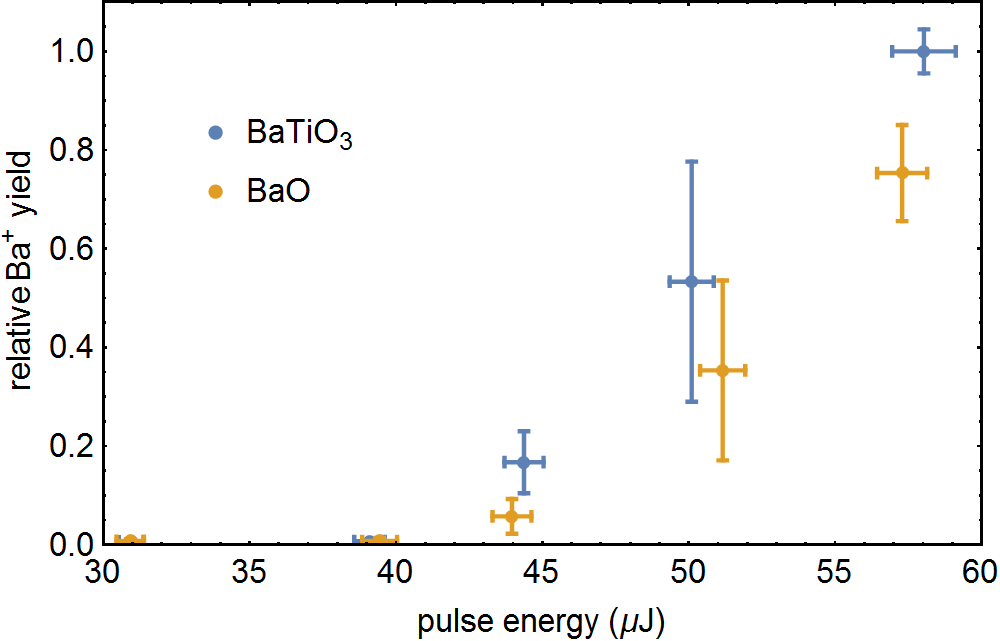}
\caption{Barium ion yield versus pulse energy.  Attenuation is added to adjust the incident pulse energy; at each pulse energy, the energies of 40 pulses are measured, and error bars represent the standard deviation of these pulse energies.  Between 3 to 7 scans are recorded at each pulse energy to determine the average relative ion yield, where a scan consists of an average of 25 TOF spectra; error bars represent the standard deviation of the relative ion yield in these scans.  Both samples indicate a threshold of about 42 $\mu$J for efficient barium ion production; for an elliptical beam with 280 $\mu$m and 50 $\mu$m waists, this corresponds to a peak fluence of approximately 0.2 J/cm${}^2$.}
\label{fig:barium_ions_vs_pulse_energy}       % Give a unique label
\end{figure}

The ion yield as a function of the incident pulse energy is determined for the BaO and BaTiO${}_3$ targets, as shown in Fig.~\ref{fig:barium_ions_vs_pulse_energy}.  At each pulse energy, 3 to 7 scans are recorded, where each scan consists of an average of 25 TOF spectra (25 ablation pulses), and the average peak height is used to determine the relative ion yield.  The pulse energy is adjusted by adding attenuation.  Using a pyroelectric energy sensor (Thorlabs, ES111C), we measure the reflectivity of M1 and the transmission of the vacuum viewport, and the pulse energy prior to M1 at each attenuation level, and thus determine the pulse energy for each attenuation level at the sample.  Both samples exhibit a threshold for efficient barium ion production near 42 $\mu$J, or a peak fluence of approximately 0.2 J/cm${}^2$.  We conclude the ablation laser beam path should be carefully designed to maintain fluence above this threshold for efficient ion trap loading.

%

%%%%%%%%%%%%%%%%%%%%%%%%%%%%%%%%%%%%%%%%%%
\section{Ion Trap Loading}
\label{sec:ion_trap}
\begin{figure}
  \includegraphics[width=1.0\columnwidth,keepaspectratio]{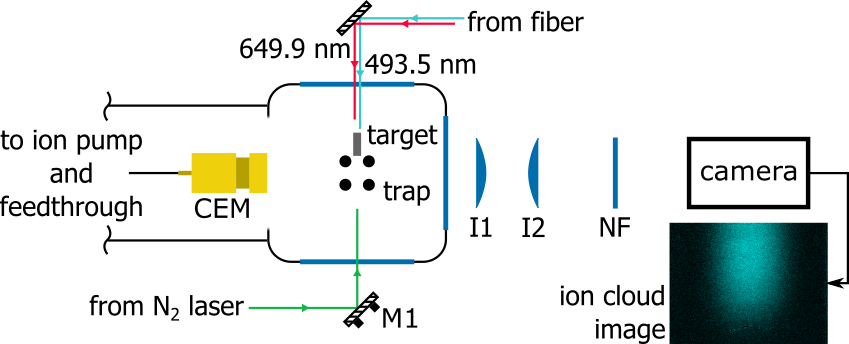}
\caption{Ion trap experiment setup (top view, schematic, not to scale).  The four rods of the trap are shown with rods protruding into the page; endcaps are omitted from the illustration.  The trap, ablation target, and CEM are housed in a stainless steel vacuum chamber at about $10^{-7}$ torr (pressure limited by chamber design and vacuum preparation, not the ablation process).  Pulses from a nitrogen laser are incident on the target to produce ions by laser ablation; the target is connected to ground.  Trapped ions are detected by either the CEM (following release from the trap) or by imaging ion fluorescence.  Light near 493.5 nm and 649.9 nm traverses the trap (above the target) to drive the ${}^2S_{1/2} \leftrightarrow {}^2P_{1/2}$ and ${}^2D_{3/2} \leftrightarrow {}^2P_{1/2}$ transitions in trapped Ba${}^+$, respectively.  Fluorescence from the trapped ions is imaged using lenses I1 and I2 onto a camera.  The inset in the lower right shows a false-color, background-subtracted image of a trapped cloud of barium ions.  CEM is channeltron electron multiplier; M1 is mirror in a piezo-actuated mount; NF is notch filter centered at 494 nm; I1 is 1-inch diameter, $f_1=75$ mm plano-convex lens; I2 is 1-inch diameter, $f_2=250$ mm plano-convex lens.  I1 is approximately $f_1$ from the trap center; I2 is approximately $f_2$ from the camera; distance between I1 and I2 is approximately 30 mm.}
\label{fig:ion_trap_setup}       % Give a unique label
\end{figure}

Barium ions produced by laser ablation of a BaTiO${}_3$ target are confined in a four-rod radiofrequency (rf) Paul trap.  The trap, chamber, and optical setup are shown schematically in Fig.~\ref{fig:ion_trap_setup}.  The trap consists of four parallel stainless steel rods with diameter of 1.6 mm and center-to-center spacing of 3.3 mm between adjacent rods (supported by macor spacers), and tungsten wire loops at each end around two opposing rods.  A helical resonator is used to apply rf voltage to two opposing rods at 11 MHz at a maximum amplitude of about 310 V, with the other two rods held at ground.  The tungsten wire loops serve as the endcaps, with a static 1 V applied to each, and an axial separation of 15 mm.  The ablation target is located about 2.5 mm axially from the trap center, is nearly flush with the outer edge of the rods, and is connected to ground.  Here, we choose BaTiO${}_3$ as the ablation target because of the higher ion yield measured previously (Fig.~\ref{fig:barium_ions_vs_pulse_number}), and because it is less brittle than the aged BaO sample.

The amplitude of the rf voltage at the output of the helical resonator and applied to the trap electrodes is determined by monitoring the voltage capacitively-coupled onto the endcap vacuum feedthrough pin.  First, a small rf calibration signal with known amplitude, directly produced by a signal generator and measured on an oscilloscope, is applied to the vacuum feedthrough pin for the trap rf electrodes.  This calibration signal is varied in amplitude and frequency to confirm the capacitively-coupled voltage at the endcap feedthrough pin is related by a constant scale factor in this range.  Then this scale factor, and the voltage measured at the endcap pin, is used to determine the rf amplitude produced by the helical resonator at the trap electrodes.  Here we assume the rf voltage at the vacuum feedthrough is equal to the rf voltage on the electrodes.

\begin{figure}
  \includegraphics[width=1.0\columnwidth,keepaspectratio]{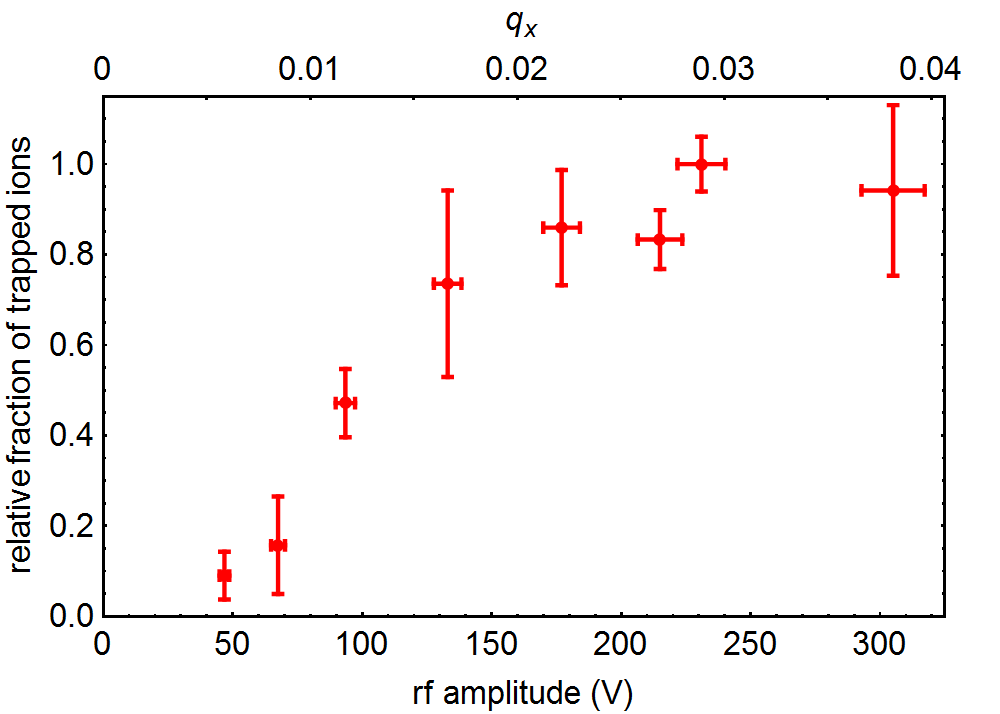}
\caption{Relative fraction of trapped ions versus rf voltage.  At each rf voltage, 3 scans are recorded to determine the average fraction of trapped ions with respect to the largest trapped fraction (at 231 V), where a scan consists of an average of 10 trap loading sequences, with the location of the ablation pulse swept across a 5x2 location grid; error bars represent the standard deviation of the average relative fraction of trapped ions in these scans.  The rf voltage is determined by the method detailed in the text; error bars represent a conservative estimate of the uncertainty in this measurement to account for possible variation in helical resonator quality factor.  Notably, the relative fraction of trapped ions plateaus above an rf voltage around 175 V.  Trapping is also parametrized in the top axis in terms of the Mathieu parameter $q_x$~\cite{champenois:ion_dynamics_tutorial} for barium ions.}
\label{fig:trapped_ions_vs_rf_voltage}       % Give a unique label
\end{figure}

We measure the relative fraction of ions trapped as a function of the applied rf voltage, with the results shown in Fig.~\ref{fig:trapped_ions_vs_rf_voltage}.  The experiment sequence is controlled by an FPGA board (Terasic, DE0-Nano).  The sequence consists of switching off the rf voltage for 1 ms, triggering the N${}_2$ laser and switching on the rf voltage, holding the trapped ions for a variable length of time (here, 10 ms), and then switching off the rf voltage to detect the ions with the CEM.  The 493.5 nm and 649.9 nm light shown in Fig.~\ref{fig:ion_trap_setup} was not present for this set of measurements.  As expected, the results shown in Fig.~\ref{fig:trapped_ions_vs_rf_voltage} indicate that the number of trapped ions decreases as the rf voltage (and thus the trap depth) approaches zero.  However, the relative fraction of ions trapped appears to plateau above an rf amplitude of about 175 V, which for an ideal linear quadrupole trap corresponds to a transverse trap depth of about 0.5 eV~\cite{champenois:ion_dynamics_tutorial}.  We conclude that above this voltage the transverse trap depth no longer limits the loading efficiency, which may indicate that the transverse trap depth is greater the average energy of the ablated ions.  However, given the dynamics of both laser ablation and the trap loading procedure, more detailed analysis would be required to eliminate all other possible limiting factors~\cite{hashimoto:ablation}.

The absolute number of ions is roughly estimated by comparing the integrated area of the CEM signal produced by a group of detected ions to that of a single detected ion.  Assuming a detection efficiency of 1-10\%~\cite{gilmore:ion_detect_efficiency_cem,sullivan:thesis_hudson_group}, we roughly estimate the average total number of ions produced by a single ablation pulse as $10^4$ to $10^5$, and the average total number of ions trapped as $10^3$ to $10^4$ (rf amplitude at 215 V).  This estimate for the number of ions produced by ablation should apply to Sec.~\ref{sec:tofms} as well, though the configuration of the TOF mass spectrometer also limits the number of ions reaching the detector.

We also observe a trapped barium ion cloud produced by laser ablation by directly imaging onto a camera the laser fluorescence from the ions, with an image shown as an inset in Fig.~\ref{fig:ion_trap_setup}.  The 493.5 nm light used to drive the ${}^2S_{1/2}$ to ${}^2P_{1/2}$ transition is produced by a custom extended-cavity diode laser (ECDL; design similar to~\cite{ricci:diode_laser}) operating near 987.1 nm that is frequency-doubled using a custom second-harmonic generation cavity (design similar to~\cite{wilson:shg,lo:shg}) with a BiBO crystal.  Light near 649.9 nm, used to drive the ${}^2D_{3/2}$ to ${}^2P_{1/2}$ transition, is directly produced by another custom ECDL.  The optogalvanic signal from a commercial, single-ended, barium hollow cathode lamp (Perkin-Elmer, N2025305) is used as a frequency reference for each laser.  A simple imaging system composed of two singlet lenses (I1, focal length $f_1=75$ mm; I2, focal length $f_2=250$ mm) and a notch filter (Semrock, FF01-494/20-25) is used to image fluorescence from trapped ions onto a camera (PointGrey, FL3-U3-13S2M-CS) with an integration time of about 333 ms, where I1 is positioned approximately $f_1$ from the ion cloud and I2 is positioned approximately $f_2$ from the camera sensor (the distance between I1 and I2 is about 30 mm).  The resulting image of the trapped ion cloud confirms the production of barium ions by laser ablation, and the applicability of this loading method for experiments with trapped ions.

%%%%%%%%%%%%%%%%%%%%%%%%%%%%%%%%%%%%%%%%%%
\section{Conclusion}
\label{sec:conclusion}
Laser ablation is a useful technique for directly producing atomic ions for trapped ion experiments.  Using a pulsed nitrogen laser, we produced Ba, Ca, Dy, Er, La, Lu, and Yb ions, and compared the relative ion yield for each.  Here, reliable production of Ba ions appears to require using substrates other than pure barium, with consistent ion production demonstrated with BaO and BaTiO${}_3$ targets, and sufficient pulsed laser fluence.  We also demonstrated loading of an rf quadrupole trap using laser ablation, and the relative loading efficiency as a function of the rf voltage.  Our results show that laser ablation may be successfully employed in future trapped ion experiments that may require or benefit from this alternative trap loading method.  Additional improvements may be gained by further characterizing the ablation process (including investigating the ablation plume~\cite{hussein:ablation_plume}), increasing the resolution of the mass spectrometer~\cite{bergmann:high_res_tof_ms_reflector,gill:laser_ablation_ion_trap_ms,schowalter:integrate_ion_trap_tof_ms}, and using resonant laser ablation~\cite{gill:resonant_laser_ablation}.

%%%%%%%%%%%%%%%%%%%%%%%%%%%%%%%%%%%%%%%%%%
\begin{acknowledgements}
We thank E. Peik for useful discussions about laser ablation; D. Hanneke for useful discussions about SHG cavity design; P. Banner, J. Hankes, and A. Nelson for technical contributions to the ion trap setup; and D. Burdick for expert machining.  P.B. acknowledges support from the J. Reid \& Polly Anderson Endowed Fund at Denison University.  This material is based upon work supported by, or in part by, the U. S. Army Research Laboratory and the U. S. Army Research Office under contract/grant number W911NF-13-1-0410; Research Corporation for Science Advancement through Cottrell College Science Award 22646; and Denison University.  Specific product citations are for the purpose of clarification only, and are not an endorsement by the authors, the U. S. Army Research Laboratory, the U. S. Army Research Office, Research Corporation for Science Advancement, or Denison University.
\end{acknowledgements}

%%%%%%%%%%%%%%%%%%%%%%%%%%%%%%%%%%%%%%%%%%
\bibliographystyle{spphys}    % APS-like style for physics

\begin{thebibliography}{10}
\bibitem{gaussian_beam_note}
Although the nitrogen laser profile shows considerable non-gaussian structure,
  we approximate it as a gaussian profile for the beam calculations presented.
\bibitem{detection_efficiency_note}
Possible mass-dependent variation in the CEM detection
  efficiency~\cite{gilmore:ion_detect_efficiency_cem} is not included in the
  ion yield calculation.
\end{thebibliography}

\end{document}